\documentclass[conference,twocolumn]{IEEEtran}

\usepackage{amsmath}
\usepackage{amsthm}
\usepackage{amssymb}
\usepackage{graphicx}
\usepackage{subfig}
\usepackage{rotating}
\usepackage{flushend}
\allowdisplaybreaks

\newsavebox{\tempbox}

\usepackage{xcolor}

\newcommand{\logt}{\log_{2}}

\newcommand{\abs}[1]{\vert #1 \vert }

\newcommand{\defn}{\triangleq}

\newcommand{\typical}[1]{T^{n}_{\left[ #1 \right] }}

\newcommand{\g}[3]{g^{n}_{ #1  }\left( #2,#3 \right)}
\newcommand{\bg}[3]{\bar{g}^{n}_{ #1  }\left( #2,#3 \right)}

\newcommand{\mcf}[1]{\mathcal{#1}}

\newcommand{\aexp}[1]{\frac{1}{n} \logt \left| {#1} \right| }

\newcommand{\cexp}[1]{\frac{1}{n} \logt {#1} }
\DeclareMathOperator{\markov}{~\begin{turn}{-41}\hspace{-5pt}$\varnothing$\end{turn}~}
\newtheorem{theorem}{Theorem}

\newtheorem{lemma}{Lemma}

\IEEEoverridecommandlockouts

\begin{document}

\title{Equating the achievable exponent region to the achievable
  entropy region by partitioning the source 
\thanks{This work was supported by the National Science Foundation
under Grant CCF-1320086.} 
}

\author{Eric Graves and Tan F. Wong \\
  Department of Electrical \& Computer Engineering \\
  University of Florida, FL 32611 \\
\texttt{\{ericsgra,twong\}@ufl.edu}
 }

\maketitle

\begin{abstract}
  In this paper we investigate the image size characterization
  problem. We show that any arbitrary source set may be decomposed
  into sets whose image size characterization is the same as its
  entropy characterization. We also show that the number of these sets
  required is small enough that one may consider that from a coding
  perspective the achievable entropy region and achievable exponent
  region are equal. This has an impact on many source networks and
  network problems whose solution heretofore could not have the image
  size characterization applied to them.
\end{abstract}

\section{Introduction}
The image-size characterization problem was originally formulated by
K{\"o}rner and Marton in \cite{KM77}, and laid some of the groundwork
for Marton's later paper on the broadcast channel~\cite{Marton79}
which established the best known inner bounds for the broadcast
channel.

Consider a discrete memoryless multiple source (DMMS) $(X^n, Y^n_{1},\dots,
Y^n_{D})$ distributed according to the joint distribution $P_{X,Y_1,
  \ldots, Y_D}$.  One can define two different but similar problems
\cite[Ch.~15]{CK}. First the \emph{entropy characterization problem}
is the categorization of the region of all tuples $(a, b_1, \ldots ,
b_{D})$ of $D+1$ non-negative numbers satisfying the condition that
for every $\delta > 0$ there exists a function $f$ with domain
$\mcf{X}^n$ such that
\begin{align*}
\left| \frac{1}{n} H(X^n | f(X^n)) - a \right| &\leq \delta \\
\left| \frac{1}{n} H(Y_i^n | f(X^n)) - b_i \right| &\leq
\delta, ~~~~~ i=1, 2, \ldots, D, 
\end{align*}
whenever $n$ is sufficiently large.  The region spawned by this
problem is referred to as $\mcf{F}$. A related region is that of
$\mcf{H}$, which is the closure of $\bigcup_{k=1}^\infty \mcf{H}_k$,
where
\begin{align*}
H_{k} &\defn
\bigg\{ \!\left( \frac{1}{k} H(X^k | U), \frac{1}{k} H(Y_1 | U),
    \ldots, \frac{1}{k} H(Y_{D}^k | U )\right) \\
& \hspace{30pt} 
:  U \markov X^k \markov Y_1^k,\ldots, Y_{D}^k \bigg\}.
\end{align*}
Interestingly $\mcf{F} = \mcf{H}$~\cite[Problem~15.17]{CK} for every DMMS; thus
characterization of both these regions may be considered as the
entropy characterization problem. A single-letter solution to the
entropy characterization problem is provided in \cite[Ch.~15]{CK} for
the special case of $3$ component sources. Only partial results
\cite[Problems~15.16--21]{CK} are available for the general case to
date.

The second problem, which will be the primary focus of this paper, is
the \emph{image characterization problem}, where we categorize the
tuples that for every $\delta>0$ and $\eta \in (0,1)$ there exists a
set $A \subseteq \typical{X}$ such that
\begin{align*}
\left| \aexp{A} - a \right| &\leq \delta \\
\left| \cexp{g^n_{Y_i|X}(A,\eta)} - b_i \right| &\leq \delta , ~~~~~ i=1, 2, \ldots, D, 
\end{align*}
whenever $n$ is large enough, where $g^n_{Y_i|X}(A,\eta)$ is the
minimum size of a set $B_i \in \mcf{Y}_i^n$ such that $P^n_{Y_i | X}(B_i
| x^n) \geq \eta$ for all $x^n \in A$. The corresponding region of
tuples is denoted as $\mcf{G}$.  Note that in the image-size
characterization problem, the conditional distributions $P_{Y_i|X},
i=1,2,\ldots, D$ describe $D$ different discrete memoryless channels.
A single-letter solution to the image-size characterization problem is
provided in \cite[Ch.~15]{CK} for the special case of $2$ component
channels.  As Csisz{\'a}r and K{\"o}rner note in \cite[pp.~339]{CK},
``image size characterizations can be used to prove strong converse
results for source networks and also to solve channel network
problems. In this respect, it is important that the sets of achievable
entropy resp. exponent triples have the same two dimensional
projections.''  It is however important to note that $\mcf{F} \neq
\mcf{G}$ in general.

Motivating this paper is that surprisingly, as shown
in~\cite[Ch.~15]{CK}, that for the case of $3$ sources (or
equivalently $2$ channels),
any triple $(a,b,c) \in \mcf{G}$ has the property that
\[
(a,b,c) = \left(\max_{i=1,2} a_i ,\max_{i=1,2}b_i,\max_{i=1,2} c_i\right),
\]
for some triples $(a_1,b_1,c_1)$ and $(a_2,b_2,c_2) \in \mcf{F}$.  The
goal of this work is to show that not only may the image sizes be
decomposed as such, but the sets themselves. Towards that goal, we
show (cf. Theorem~\ref{thm:partition}) there exists a partition of $A
= \bigcup A_k$ of no more than $n\Gamma$ component sets, for some
constant $\Gamma$, that the image size characterization problem
restricting to each of these component sets is the same as the entropy
characterization problem.  To accomplish our goal we will use
techniques associated with information spectrum~\cite{han2003}.
We first present a new method of partitioning the
destination sequence space based on information spectrum in
section~\ref{sec:lemmas}. Then we employ some consequences of such
partitioning to show the main image size characterization results in
section~\ref{sec:mt}, and finish with a remark about the significance of these results in section~\ref{sec:conc}.

\section{Notation}\label{sec:notation}
To simplify writing, let $[i:j]$ denote the set of integers starting at
$i$ and ending at $j$, inclusively.  Consider a pair of discrete
random variables $X$ and $Y$ over alphabets $\mcf{X}$ and $\mcf{Y}$,
respectively.  A set $B \subseteq \mcf{Y}^n$ is called an
\emph{$\eta$-image} of $A \subseteq \mcf{X}^n$ over the channel
$P^n_{Y|X}$ \cite[Ch. 15]{CK} if $P^n_{Y|X}(B| x^n) \geq \eta$ for
every $x^n \in A$. On the other hand, $B$ is called an
\emph{$\eta$-quasi-image} of $A$ over the channel $P^n_{Y|X}$
\cite[Problem~15.13]{CK} if $\Pr\{ Y^n \in B | X^n \in A\} \geq \eta$.
The minimum size of $\eta$-images of $A$ over $P^n_{Y|X}$ will be
denoted by $g^n_{Y|X}(A,\eta)$, while the minimum size of
$\eta$-quasi-images of $A$ over $P^n_{Y|X}$ will be denoted by $\bar
g^n_{Y|X}(A,\eta)$.

The notation for conditional entropy will be slightly abused
throughout the paper. Within, when a quantity such as $H(Y^n | X^n \in
A')$ is expressed it will mean $H(Y^n | E=1)$, where $E$ is a random
variable taking the value $1$ if $X^n \in A'$ and $0$ if not. This is
in contrast to the proper use in which $H(Y^n | X^n \in A')$ would
equal $H(Y^n | E)$.

\section{Information Spectrum Partition}\label{sec:lemmas}

Let $p_n$ be a distribution on $\mcf{Y}^n$, and $i_n = -\cexp{p_n}$ be
the corresponding information spectrum. For any $\delta > 0$, 
define $K_{\delta} \defn \left\lceil \frac{\logt
    \abs{\mcf{Y}}}{\delta} \right\rceil$ and the $\delta$-information
spectrum partition of $\mcf{Y}^n$ with respect to (w.r.t.)  $i_n$ be
$\{B_k\}_{k=0}^{K_{\delta}}$, where
\[
B_k \defn 
\begin{cases}
\{ y^n \in \mcf{Y}^n : k\delta\leq i_n(y^n) <
(k+1)\delta\} \\
& \hspace{-80pt} \text{ for } k\in [0:K_{\delta} -1] \\
\{ y^n \in \mcf{Y}^n :  K_{n}\delta\leq i_n(y^n) < \infty \} & \\
&\hspace{-80pt} \text{ for } k = K_{\delta}.
\end{cases}
\]
For convenience, we sometimes associate $B_{-1} \defn \emptyset$ and
the zero-probability set $B_{\infty} \defn \{ y^n : p_n(y_n) = 0 \}$
with $\{B_k\}$. Clearly $\mcf{Y}^n = \left(\bigcup_{k=0}^{K_{\delta}}
  B_k \right) \cup\ B_{\infty} $.

Fix any $\delta \in (0,1)$ and $A \subseteq \mcf{X}^n$ with the
property that $\Pr\{ X^n \in A \} > 0$.  For the rest of this section,
let $P_A(y^n) \defn \Pr\{ Y^n = y^n | X^n \in A\}$ and $\{B_k\}$ be
the $\delta$-information spectrum partition of $\mcf{Y}^n$
w.r.t. $-\cexp{P_A}$. We may derive the following consequences of
information partitions. 

\begin{lemma}\label{lem:bk_size}
For every $B_{k}$, $k \in [0:K_{\delta}]$,
\[
  \aexp{B_k} < (k+1) \delta. 
\]
In addition if $P_A(B_k) > 2^{-n \delta}$ then
\[
 \left| \aexp{B_k} - k \delta \right| < \delta.
\]
\end{lemma}
\begin{IEEEproof}
Trivially we have 
\[
\aexp{B_{K_{\delta}}}  \leq \logt\abs{\mcf{Y}} < (K_{\delta}+ 1)  \delta.
\]
For $k \in [0:K_{\delta} - 1]$,
 \[
 1 \geq \sum_{y^n \in B_{k}} P_A(y^n) > |B_{k}| 2^{-n (k+1) \delta},
 \]
 and therefore $\aexp{B_k} < (k+1) \delta$.  Similarly suppose that
 $P_A(B_k) = \sum_{y^n \in B_k} P_A(y^n) > 2^{-n \delta}$. Then
 \[
 2^{-n \delta} < \sum_{y^n \in B_{k} } P_A(y^n) \leq 2^{-n k
   \delta} | B_k|,
 \]
 and therefore $\aexp{B_k} > (k-1)\delta$. Combining both results
 gives us $ \left| \aexp{B_k} - k \delta \right| < \delta$.
 \end{IEEEproof}

\begin{lemma}\label{lem:barg_bound}
  For any $\eta \in (0,1]$ and sufficiently large $n$, there exists a
  $k' \in [0:K_{\delta}]$ such that
\[
\cexp{\bar g^n_{Y|X}(A,\eta) } \leq 
\aexp{ \bigcup_{k=0}^{k'} B_{k} }
\leq (k'+2) \delta.
\]
Furthermore if $P_A(B_{k'-1}) > 2^{-n \delta}$, then
\[
\cexp{\bar g^n_{Y|X}(A,\eta) } \geq \aexp{B_{k'-1}} \geq (k'-2)
\delta.
\]
\end{lemma}
\begin{IEEEproof}
  Let $\eta_{-1} \defn 0$ and $\eta_{k'} \defn P_A
  \left(\bigcup_{k=0}^{k'} B_k \right)$ for $k' \in [0: K_{n}]$. Note
  then that $0=\eta_{-1} \leq \eta_0 \leq \eta_1 \leq \cdots \leq
  \eta_{K_{\delta}} = 1$. In addition $\eta_{k-1} = \eta_k$ implies
  $B_k = \emptyset$. Write $B' = \bigcup_{k=0}^{k'} B_k$ to simplify
  notation below. Clearly $B'$ is an $\eta_{k'}$-quasi-image of
  $A$. We claim that $B'$ is in fact the unique
  $\eta_{k'}$-quasi-image of $A$ that achieves the minimum size $\bar
  g^n_{Y|X}(A,\eta_{k'})$.  To show the claim, consider a set $\hat B
  \subseteq \mcf{Y}^n$ such that $\hat B \neq B'$ and $|\hat B| \leq
  |B'|$. For $k'=K_{\delta}$, $\hat B$ clearly cannot be the
  $\eta_{k'}$-quasi-image of $A$. On the other hand, for $k' \in
  [0:K_{\delta}-1]$,
\begin{align*}
&P_A(\hat B)
 = P_A(B') -  P_A(B' \setminus \hat B) + P_A(\hat B \setminus B') \\
& =  \eta_{k'} - \sum_{k=0}^{k'} \sum_{y^n \in B_k \setminus \hat B}
P_A(y^n) 
+ \sum_{k \geq k'+ 1} \sum_{y^n \in B_k \cap \hat B} P_A(y^n) \\
&<  \eta_{k'} 
- \left( |B'| - |B' \cap \hat B| \right) 2^{-n(k'+1) \delta} \\
& ~~~~+ \left( |\hat B| - |B' \cap \hat B| \right) 2^{-n(k'+1) \delta}  
\leq \eta_{k'}.
\end{align*}
Thus $\hat B$ cannot be an $\eta_{k'}$-quasi-image of $A$. 

Next it is clear that for any $\eta \in (0,1]$ there exists a $k'$ such that
$ \eta_{k'-1} < \eta \leq \eta_{k'}$ which gives us that
\begin{equation}
\aexp{ \bigcup_{k=0}^{k'-1} B_{k} } \leq \cexp{\bar g^n_{Y|X}(A,\eta) }
 \leq \aexp{ \bigcup_{k=0}^{k'} B_{k} }.
\label{eq:lem3_1}
\end{equation}
Lemma~\ref{lem:bk_size} and the upper bound in \eqref{eq:lem3_1} give
us that
\[
\cexp{\bar g^n_{Y|X}(A,\eta) } 
\leq \cexp{ \sum_{k=0}^{k'} 2^{n(k+1) \delta} }
\leq (k'+2) \delta
\]
when $n$ is sufficiently large.  Furthermore if $P_A(B_{k'-1}) > 2^{-n
  \delta}$, then combining the lower bound of \eqref{eq:lem3_1} and
Lemma~\ref{lem:bk_size} again we have
\[
\cexp{\bar g^n_{Y|X}(A,\eta) } \geq \aexp{B_{k'-1}} \geq (k'-2) \delta.
\]
\end{IEEEproof}

\begin{lemma}\label{lem:6.6}
  Fix any $\alpha' \in (0,1)$. For any 
  $\alpha_n \in (0,\alpha']$ with $\frac{-\logt \alpha_n}{n}
  \rightarrow 0$, there exist $\beta_n \rightarrow 1$ and $\tau_n
  \rightarrow 0$ such that
\[
0 \leq \cexp{ \g{Y|X}{A'}{\beta_n}} - \cexp{ \g{Y|X}{A'}{\alpha_n}}
\leq \tau_n
\]
for every $A' \subseteq \mcf{X}^n$, whenever $n$ is sufficiently
large. Furthermore the same $\beta_n$ and $\tau_n$ can be used
uniformly for all $\alpha_n \geq \frac{1}{n^2}$.
\end{lemma}
\begin{IEEEproof}
  This is a slightly strengthened version of \cite[Lemma~6.6]{CK},
  whose proof (cf. also \cite[Ch.~5]{CK} ) directly applies to the
  current lemma.
\end{IEEEproof}

\begin{lemma} \label{lem:ApsubA} 
 
  Let $X^n$ be conditionally uniformly distributed over $A$.  Then for
  any $\alpha_n \in (0,1]$ with $\frac{-\logt \alpha_n}{n} \rightarrow
  0$, 
  there exist $A' \subseteq A$, $\tau_n \rightarrow 0$, and $\beta_n
  \rightarrow 1$ such that 
  $\frac{\abs{A'}}{\abs{A}} \geq \left(1-\frac{1}{n} \right) \alpha_n$ and
  \[
  \cexp{\g{Y|X}{A'}{\beta_n}} \leq \cexp{\bg{Y|X}{A}{\alpha_n}} +
  \tau_n,
  \]
  whenever $n$ is sufficiently large. Neither $\tau_n$ nor $\beta_n$
  depends on $A$. Furthermore neither depends on $\alpha_n$ if
  $\alpha_n \geq \frac{1}{n}$.
\end{lemma}
\begin{IEEEproof}
  Let $B\subseteq \mcf{Y}^n$ be an $\alpha_n$-quasi-image of $A$ that
  achieves $\bg{Y|X}{A}{\alpha_n}$. 
Define
\[
A' \defn \left\{x^n \in A: P^n_{Y|X}(B | x^n) \geq
\frac{\alpha_n}{n} \right\}.
\]
Clearly $B$ is an 
$\frac{\alpha_n}{n}$-image of $A'$. Hence
\begin{align*}
\cexp{\bg{Y|X}{A}{\alpha_n}} &=\aexp{B} \\
&\geq 
\cexp{\g{Y|X}{A'}{\frac{\alpha_n}{n}}} \\
&\geq 
\cexp{\g{Y|X}{A'}{\beta_n}} - \tau_n
\end{align*}
by Lemma~\ref{lem:6.6} since 
$\frac{\logt n -\logt \alpha_n}{n} \rightarrow 0$.  Note that the same
$\beta_n$ and $\tau_n$ can be used uniformly for all 
$\alpha_n \geq \frac{1}{n}$.

Further as $B$ is an $\alpha_n$-quasi-image of $A$, we have
\begin{align*}
\alpha_n &\leq P_A(B) 
\\
&= \frac{1}{\abs{A}} \sum_{x^n \in A} P^n_{Y|X} (B|x^n) \\
&= \frac{1}{\abs{A}} \sum_{x^n \in A'} P^n_{Y|X} (B| x^n) +
\frac{1}{\abs{A}} \sum_{x^n \in A \setminus A'} P^n_{Y|X}(B| x^n)  
\\
&\leq  \frac{\abs{A'}}{\abs{A}} 
+  \left(1- \frac{\abs{A'}}{\abs{A}} \right) 
\frac{\alpha_n}{n}
\end{align*}
which implies 
$\frac{\abs{A'}}{\abs{A}} \geq (1-\frac{1}{n}) \alpha_n$.
\end{IEEEproof}

\begin{lemma}\label{lem:daco}
  Suppose that $X^n$ is conditionally uniformly distributed on $A$.
  Then there exist $A^\ast \subseteq A$, $\varepsilon_n \rightarrow
  0$, and $\beta_n \rightarrow 1$ satisfying
  $\frac{\abs{A^\ast}}{\abs{A}} \geq \frac{1}{2(K_{\delta}+1)}$ and
\[
\frac{1}{n}H(Y^n | X^n \in A^\ast) \geq \cexp{g^n_{Y|X}(A^\ast,\beta_n)} 
- 7.19 \delta - \varepsilon_n,
\]
whenever $n$ is sufficiently large. Neither $\varepsilon_n$ nor
$\beta_n$ depends on $A$.
\end{lemma}

\begin{IEEEproof}
  Define $\eta_{k} \defn P_A \left( \bigcup_{l=0}^{k} B_l \right)$ for
  $k\in [0:K_{n}]$ as in the proof of Lemma~\ref{lem:barg_bound}.
  Because the total number of sets in $\{B_k\}$ is $K_{\delta}+1$, we
  know that there exists at least one $k' \in [0,K_{\delta}]$ such
  that $P_A(B_{k'}) \geq \frac{1}{K_{\delta}+1}$. Apply
  Lemma~\ref{lem:ApsubA} by choosing $\alpha_{n} = \eta_{k'} \geq
  \frac{1}{2(K_{\delta}+1)}$ 
  to obtain $\tau_{n} \rightarrow 0$, $\beta_{n} \rightarrow 1$, and
  $A' \subseteq A$ that satisfy
\begin{align}  
\frac{\abs{A'}}{\abs{A}} 
&\geq 
\left(1-\frac{1}{n} \right)\eta_{k'}
\geq \frac{1}{2(K_{\delta}+1)},
\label{eq:A/A'} \\
\cexp{\g{Y|X}{A'}{\beta_{n}}} &\leq
\cexp{\bg{Y|X}{A}{\eta_{k'}}} + \tau_{n}, 
\label{eq:g<bg}
\end{align}
whenever $n$ is sufficiently large. Note that the $\beta_n$ and
$\tau_n$ above are the ones that work uniformly for all $\alpha_n \geq
\frac{1}{n}$ in Lemma~\ref{lem:ApsubA}.

First consider the case of $k' \leq c_n \defn 4.19 +
\frac{\tau_{n}}{\delta}$.  From \eqref{eq:g<bg},
\begin{align}
& \hspace*{-10pt} \cexp{\g{Y|X}{A'}{\beta_{n}}} 
\leq \cexp{\bg{Y|X}{A}{\eta_{k'}} } + \tau_{n} \notag \\
& \stackrel{(a)}{=}  \aexp{ \bigcup_{k =0}^{k'}  B_{k}}  + \tau_{n} 
\leq (k'+2)\delta + \tau_{n} \label{e:bb} \\
&\leq 6.19 \delta + 2\tau_{n}. \notag
\end{align}
where (a) is due to the fact that $\bigcup_{k =0}^{k'} B_{k}$ is the
$\eta_{k'}$-quasi-image of $A$ that achieves
$\bg{Y|X}{A}{\eta_{k'}}$ as shown in the proof of
Lemma~\ref{lem:barg_bound}. Since $H(Y^n | X^n \in A') \geq 0$,
the conclusions of the lemma are clearly satisfied.

It remains to consider the case of $k' > c_n$. To that end, let $k''
\defn \lfloor k'-c_n \rfloor$, and define set $\tilde B =
\bigcup_{k=0}^{k''} B_{k}$. First assume that $P_A(\tilde B) =
\eta_{k''} > \frac{1}{n}$. Apply Lemma~\ref{lem:ApsubA} again with
$\alpha_{n} = \eta_{k''} > \frac{1}{n}$ 
to obtain $A'' \subseteq A$ that satisfies
\begin{align}
& \hspace*{-10pt}
\cexp{g^n_{Y|X}(A'',\beta_{n})} \leq 
\cexp{\bar g^n_{Y|X}(A,\frac{1}{n} )} + \tau_{n} \notag \\
& \leq \cexp{\bar g^n_{Y|X}(A,\eta_{k''} )} + \tau_{n} 
\stackrel{(a)}{\leq} (k'' + 2)\delta  + \tau_{n} \notag \\
&\leq (k' -2.19)\delta 
\stackrel{(b)}{\leq} \aexp{B_{k'}} -1.19 \delta
\label{eq:bkpmg}
\end{align}
where (a) and (b) are due to Lemmas~\ref{lem:barg_bound} and
\ref{lem:bk_size}, respectively.

Let $\hat B$ be the $\beta_{n}$-image of $A''$ that achieves
$\g{Y|X}{A''}{\beta_{n}}$. By definition, every $y^n \in B_{k'}$ has
the property that $2^{-n(k'+1) \delta} < P_A(y^n) \leq 2^{-n k'
  \delta}$. This implies
\begin{align*}
& \hspace*{-10pt} P_A(B_{k'} \setminus \hat B) 
= P_A(B_{k'})  - P_A(B_{k'} \cap \hat B)  \\
&\geq \frac{1}{K_{\delta}+1} -   2^{-n k' \delta} \left| B_{k'} \cap \hat B \right| 
\\
&\geq \frac{1}{K_{\delta}+1} -  2^{-n k' \delta} g^n_{Y|X}(A'',\beta_{n})  \\
&\geq  \frac{1}{K_{\delta}+1} - 2^{n \delta}
\frac{\g{Y|X}{A''}{\beta_{n}}}{\left| B_{k'} \right|} 
\\
& \geq  \frac{1}{K_{\delta}+1} - 2^{-0.19n\delta}
\end{align*}
where the second last and last inequalities are due to
Lemma~\ref{lem:bk_size} and \eqref{eq:bkpmg}, respectively. 
Continuing on,
\begin{align*}
& \hspace*{-10pt} 
\frac{1}{K_{\delta}+1} - 2^{-0.19n\delta} 
\leq P_A (B_{k'} \setminus \hat B)  \\
&= \frac{1}{|A|} \sum_{x^n \in A} P^n_{Y|X}(B_{k'} \setminus \hat B | x^n)\\
&\stackrel{(a)}{=}
\frac{1}{|A|} \sum_{x^n \in A'' } P^n_{Y|X}(B_{k'} \setminus \hat B | x^n ) \\
&\hspace{10pt}+ \frac{1}{|A|} \sum_{x^n \in A' \setminus A'' } 
P^n_{Y|X}( B_{k'} \setminus \hat B | x^n ) \\
&\hspace{10pt}+ \frac{1}{|A|} \sum_{x^n \in A \setminus (A' \cup A'')  } 
\hspace{-10pt} P^n_{Y|X}(B_{k'} \setminus \hat B | x^n) \\
&\stackrel{(b)}{\leq}
(1-\beta_{n})  + \frac{\abs{A' \setminus A'' }}{\abs{A}}
+\frac{\eta_{k'}}{n}
\end{align*}
where each term in (b) bounds the corresponding term in (a). In
particular, the first bound in (b) is due to the fact that each $x^n
\in A''$ satisfies $P^n_{Y|X}(\hat B^c | x^n) < 1- \beta_{n}$. On
the other hand, the third bound in (b) results from the fact that $A'$
contains all $x^n \in A$ that $P^n_{Y|X} \left( \bigcup_{k=0}^{k'}
  B_{k} | x^n \right) \geq \frac{\eta_{k'}}{n}$ as defined in the
proof of Lemma~\ref{lem:ApsubA} because $\bigcup_{k=0}^{k'} B_{k}$ is
the unique minimum-cardinality $\eta_{k'}$-quasi-image of $A$ (cf. the
proof of Lemma~\ref{lem:barg_bound}).  As a result, we have
\begin{align}
\frac{\abs{A' \setminus A''}}{\abs{A}}
& \geq 
\frac{1}{K_{\delta}+1} - 2^{-0.19n\delta}
-(1- \beta_{n}) - \frac{1}{n} \notag \\ & 
\geq \frac{1}{2(K_{\delta}+1)}
\label{eq:Aminus}
\end{align}
for all sufficiently large $n$.  Now since $X^n$ is conditionally
uniform in $A$, we have
\begin{align}
& \hspace*{-5pt} 
P_{A'\setminus A''}(y^n) \defn
\Pr( Y^n =y^n | X^n \in A'\setminus A'') \notag \\
&= \frac{1}{\left| A' \setminus A'' \right|} \sum_{x^n \in A' \setminus A'' }
P^n_{Y|X}(y^n | x^n) \notag \\
&\leq \frac{2(K_{\delta}+1)}{|A|} \sum_{x^n \in A} P^n_{Y|X}(y^n | x^n)
\leq 2(K_{\delta}+1) P_{A}(y^n). \label{eq:ptyy}
\end{align}
Hence using \eqref{eq:ptyy} we get
\begin{align}
& \hspace*{-1pt} \frac{1}{n} H(Y^n | X^n \in A'\setminus A'') 
\notag \\
&\geq - \frac{1}{n} \sum_{y^n \notin \tilde B } P_{A'\setminus A''}(y^n) 
 \logt P_{A'\setminus A''}(y^n) \notag \\
&\geq -\frac{\logt 2(K_{\delta}+1)}{n}  - \frac{1}{n}
\sum_{k=k''}^{K_{\delta}} \sum_{y^n \in B_k} P_{A'\setminus A''}(y^n) \logt
P_{A}(y^n)  \notag \\
&\geq -\frac{\logt 2(K_{\delta}+1)}{n} + \sum_{k=k''}^{K_{\delta}} 
P_{A'\setminus A''}(B_{k}) \cdot k \delta \notag \\
&\geq -\frac{\logt 2(K_{\delta}+1)}{n} + (k'-c_n-1)\delta P_{A'\setminus A''}(\tilde B^c)
\notag \\
& \stackrel{(a)}{\geq}
  -\frac{\logt 2(K_{\delta}+1)}{n} + \bigg( \cexp{\g{Y|X}{A'}{\beta_{n}}} -
  7.19\delta \notag \\
& \hspace*{60pt} - 2\tau_{n} \bigg) \cdot P_{A'\setminus A''}(\tilde B^c) \notag \\
& \stackrel{(b)}{\geq}
-\frac{\logt 2(K_{\delta}+1)}{n} + \bigg( \cexp{\g{Y|X}{A' \setminus A''}{\beta_{n}}} 
 \notag \\
& \hspace*{60pt} 
- 7.19\delta - 2\tau_{n} \bigg) 
\cdot \left( 1 - \frac{\eta_{k''}}{n}  \right)  \notag \\
& \geq \cexp{\g{Y|X}{A' \setminus A''}{\beta_{n}}} -\frac{\logt
  \left[2(K_{\delta}+1) \logt \abs{\mcf{Y}}\right]}{n}  \notag \\
& \hspace*{10pt} - 7.19\delta - 2\tau_{n}
\label{eq:hty2}
\end{align}
where (a) is due to \eqref{e:bb} and (b) is due to the fact that
$A''$ contains all $x^n \in A$ that $P^n_{Y|X} \left( \tilde B|
  x^n \right) \geq \frac{\eta_{k''}}{n}$.  Clearly then the
conclusions of the lemma result from \eqref{eq:Aminus} and
\eqref{eq:hty2}.

Finally if $P_A(\tilde B) \leq \frac{1}{n}$, then 
following the same development from \eqref{eq:Aminus} to
\eqref{eq:hty2} based on \eqref{eq:A/A'}, we get $P_{A'}(y^n) \leq
2(K_{\delta}+1)P_{A}(y^n)$ and
\begin{align*}
& \hspace*{-10pt} \frac{1}{n} H(Y^n | X^n \in A') \\
& \geq -\frac{\logt 2(K_{\delta}+1)}{n} + \bigg( \cexp{\g{Y|X}{A'}{\beta_{n}}} 
  \\
& \hspace{40pt} - 7.19\delta -2\tau_{n} \bigg)\cdot \left( 1-
  2(K_{\delta}+1)P_{A}(\tilde B) \right) \\
& \geq \cexp{\g{Y|X}{A'}{\beta_{n}}} -\frac{\logt 2(K_{\delta}+1)}{n}
\\
& \hspace*{10pt} - \frac{2 (K_{\delta}+1) \logt \abs{\mcf{Y}}}{n}  - 7.19\delta -2\tau_{n} .
\end{align*}
This, together with \eqref{eq:A/A'}, again gives the lemma.
\end{IEEEproof}

\section{Image Size Characterization}\label{sec:mt}

\begin{theorem}\label{thm:main}
  Fix any $\eta\in (0,1)$ and $\epsilon > 0$. Let $X^n$ be uniformly
  distributed over any $A \subseteq \mcf{X}^n$. For $i \in [1:D]$,
  suppose that $Y^n_i$ is conditionally distributed according to the
  channel $P^n_{Y_i|X}$ given $X^n$.  Then there exists $A'\subseteq
  A$ satisfying
\begin{enumerate}
\item $0 \leq \aexp{A} - \aexp{A'} \leq \epsilon$,
\item $\frac{1}{n}H(X^n | X^n \in A') =  \aexp{A'}$, and
\item $\left|\frac{1}{n}H(Y_i^n| X^n \in A')
    -\cexp{g^n_{Y_i|X}(A',\eta) }\right| \leq \epsilon$ for $i \in
  [1:D]$,
\end{enumerate}
whenever $n$ is sufficiently large.
\end{theorem}

\begin{IEEEproof}
  We give the proof for the cases of $D=1$ and $2$ below. The proof
  naturally extends for $D>2$.

  Apply Lemma~\ref{lem:daco} based on the $\delta_1$-information
  spectrum partition of $\mcf{Y}_1^n$ to obtain $A_1 \subseteq A$ and
  $\varepsilon_{n} \rightarrow 0$ such that
\begin{align}
& \aexp{A} - \aexp{A_1}  \leq  \cexp{2 (K_{\delta_1}+1)}
\label{eq:A1} \\
&\cexp{\g{Y_1|X}{A_1}{\eta}} 
  \leq \frac{1}{n}H(Y_1^n | X^n \in A_1) + \varepsilon_{n} +
  7.19\delta_1,
\label{eq:gA1<}
\end{align}
for all sufficiently large $n$. On the other hand, for all
sufficiently large $n$, by \cite[Lemma~15.2]{CK}
\begin{equation}
\frac{1}{n}H(Y_1^n | X^n \in A_1) \leq
\cexp{\g{Y_1|X}{A_1}{\eta}} + \epsilon.
\label{eq:gA1>}
\end{equation}
Note that \eqref{eq:A1}, \eqref{eq:gA1<}, and \eqref{eq:gA1>} together
with a small enough $\delta_1$ establish the theorem for the case of
$D=1$.

Next apply Lemma~\ref{lem:daco} based on the $\delta_2$-information
spectrum partition of $\mcf{Y}_2^n$ and \cite[Lemma~15.2]{CK} again to
obtain $A_2 \subseteq A_1$ such that
\begin{align}
& \aexp{A_1} - \aexp{A_2}  \leq  -\cexp{2 (K_{\delta_2}+1)}, 
\label{eq:A2} \\
&\cexp{\g{Y_2|X}{A_2}{\eta}} \leq 
\frac{1}{n}H(Y_2^n | X^n \in A_2) + \varepsilon_n + 7.19\delta_2,
\label{eq:gA2<} \\
&\frac{1}{n}H(Y_2^n | X^n \in A_2) \leq
\cexp{\g{Y_2|X}{A_2}{\eta}} + \epsilon.
\label{eq:gA2>}
\end{align}
whenever $n$ is sufficiently large. Furthermore, applying
\cite[Lemma~15.2]{CK} on $A_2$ and $A_1\setminus A_2$ over the first
channel gives us, respectively,
\begin{align}
&\hspace*{-10pt} \frac{1}{n}H(Y_1^n | X^n \in A_2) \leq
\cexp{\g{Y_1|X}{A_2}{\eta}} + \epsilon, 
\label{eq:gA12>} \\
& \hspace*{-10pt} \frac{1}{n}H(Y_1^n | X^n \in A_1 \setminus A_2) \notag \\
&\leq
\cexp{\g{Y_1|X}{A_1\setminus A_2}{\eta}} + \frac{\epsilon}{4
  (K_{\delta_2}+1)} \notag \\
& \leq
\frac{1}{n}H(Y_1^n | X^n \in A_1) + \varepsilon_{n} + 7.19\delta_1 +
\frac{\epsilon}{4 (K_{\delta_2}+1)} 
\label{eq:H1-2}
\end{align}
where the last inequality is due to \eqref{eq:gA1<}.

Now let $S$ be in the indicator random variable of the
event that $X^n \in A_2$. We have
\begin{align}
& \hspace*{-10pt} H(Y_1^n | X^n \in A_1)  \notag \\
&=  I(S ; Y_1^n | X^n \in A_1) + H(Y_1^n | S, X^n \in A_1) \notag\\
& \leq 1+H(Y_1^n | X^n \in A_2) \Pr\{X^n \in A_2 |X^n \in A_1\} \notag
\\
& \hspace{10pt}
+ H(Y_1^n | X^n \in A_1 \setminus A_2)\Pr\{ X^n \in A_1
  \setminus A_2 |X^n \in A_1\}\notag \\
& \leq 1 + H(Y_1^n | X^n \in A_2) \cdot
\frac{\abs{A_2}}{\abs{A_1}} + \bigg[ H(Y_1^n | X^n \in A_1) \notag \\
& \hspace{30pt}
+ n\varepsilon_n + 7.19n\delta_1 + \frac{n\epsilon}{4 (K_{\delta_2}+1)} \bigg] \cdot
 \left[ 1 -  \frac{\abs{A_2}}{\abs{A_1}}\right]
\label{eq:thm1_bound}
\end{align}
where the last inequality is due to \eqref{eq:H1-2}.  Because of
\eqref{eq:A2}, we can rearrange \eqref{eq:thm1_bound} to get
\begin{align}
& \hspace*{-10pt} \frac{1}{n}  H(Y_1^n | X^n \in A_2)  \notag \\
&\geq
\frac{1}{n}  H(Y_1^n | X^n \in A_1) + 7.19\delta_1 + \varepsilon_n
\notag \\ 
& \hspace{10pt} - \left(\varepsilon_n + 7.19\delta_1
  + \frac{\epsilon}{4(K_{\delta_2}+1)}  + \frac{1}{n}\right) \frac{\abs{A_1}}{\abs{A_2}}
\notag \\
&\geq
\cexp{\g{Y_1|X}{A_2}{\eta}} - \frac{\epsilon}{2}  \notag \\
& \hspace{10pt} - 2(K_{\delta_2}+1)\left(\varepsilon_n + 7.19\delta_1 + \frac{1}{n} \right) 
\label{eq:gA12<}
\end{align}
where the last inequality is due to \eqref{eq:gA1<} and
\eqref{eq:A2}. Finally putting \eqref{eq:A1}, \eqref{eq:A2},
\eqref{eq:gA2<}, \eqref{eq:gA2>}, \eqref{eq:gA12>}, and
\eqref{eq:gA12<} together with small enough $\delta_1$, $\delta_2$,
and $\frac{\delta_1}{\delta_2}$, we get the theorem for the case of
$D=2$.
\end{IEEEproof}

\begin{theorem} \label{thm:partition}
  Fix any $\eta\in (0,1)$ and $\epsilon > 0$.  Let $X^n$ be uniformly
  distributed over any $A \subseteq \mcf{X}^n$. For $i \in [1:D]$,
  suppose that $Y^n_i$ is conditionally distributed according to the
  channel $P^n_{Y_i|X}$ given $X^n$.  Then there exist a constant
  $\Gamma >0 $ and a partition of $A = \bigcup_{k=1}^{m} A_k$ with $m
  \leq n\Gamma$ that satisfies
\begin{enumerate}
\item $\frac{1}{n}H(X^n | X^n \in A_k) =  \aexp{A_k}$ and
\item $\left|\frac{1}{n}H(Y_i^n| X^n \in A_k)
    -\cexp{g^n_{Y_i|X}(A_k,\eta) }\right| \leq \epsilon$ for $i \in
  [1:D]$,
\end{enumerate}
for all $k \in [1:m]$, whenever $n$ is sufficiently large.
\end{theorem}
\begin{IEEEproof}
  Using Theorem~\ref{thm:main} on $A$, we immediately obtain $A_1
  \subseteq A$ that satisfies 1) and 2). In addition, $\Pr\{ X^n \in A
  \setminus A_1\} \leq \delta$ for some $\delta \in (0,1)$. Next apply
  Theorem~\ref{thm:main} again on $A\setminus A_1$, we get $A_2
  \subseteq A\setminus A_1$ satisfying 1), 2), and $\Pr \{ X^n \in A
  \setminus (A_1 \cup A_2) | X^n \in A \setminus A_1\} \leq \delta$.
  Repeat this
  process $m-2$ more times to get $A_k \subseteq A \setminus
  \bigcup_{j=1}^{k-1} A_j$ satisfying 1), 2), and
\[
  \Pr \bigg\{ X^n \in A\setminus \bigcup_{j=1}^{k} A_j ~\bigg|~ X^n \in A\setminus
  \bigcup_{j=1}^{k-1} A_j \bigg\}\leq \delta
\]
for
  $k \in [3:m]$. Write $\tilde A \defn \bigcup_{j=1}^{m} A_j$. Then
  combining the conditional probability bounds above, we have
  $\Pr\{X^n \in \tilde A \} \leq  \delta^m$. Since $X^n$ distributed
  uniformly in $A$, $\Pr\{X^n \in \tilde A \} \geq
  2^{-n\logt\abs{\mcf{X}}}$. Thus $\tilde A$ must be empty when $m
 > \frac{n}{-\log_{\abs{\mcf{X}}} \delta}$.
\end{IEEEproof}

\section{Concluding Remark} \label{sec:conc}
Consider a coding application in which the set $A \subseteq \mcf{X}^n$
represents the codebook. Theorem~\ref{thm:partition} tells us that $A$
can be broken down into $\bigcup_{k=1}^{m} A_k$ of at most $n \Gamma$
sets. Let $E =k$ if $X^n \in A_k$ for $k \in [1:m]$, and hence $H(E)
\leq \logt n +\logt \Gamma $. For any message $M$ of rate $R$ carried
by the codebook $A$ to be received at receiver $i$, by Fano's Inequality we have
\begin{align*}
R &\leq \frac{1}{n}I(M;Y_i^n) + \frac{1}{n} + P_e R \\
&\leq I(M;Y_i^n | E) +\frac{1+\logt n}{n} + P_e R
\end{align*}
where $P_e$ is the error probability of decoding $M$ based on
observing $Y_i^n$. Thus it suffices to restrict to those codewords
with each $A_k$. Within each $A_k$ the image-size characterization,
which is important to further bounding $I(M;Y_i^n | E=k)$, is the same
as the achievable entropy characterization.  Conversely, suppose that
for each $k$, one has a coding scheme to send $M$, restricted to
$A_k$, through the $i$th channel that achieves the rate
$\frac{1}{n}I(M;Y_i^n |E=k)$. Then one can derive a scheme to first
send $E$ and then send $M$ to achieve rate $\frac{1}{n} I(M;Y_i^n | E)$
because the number of bits needed to communicate $E$ is negligible
compared to that required to send $M$. Once again the achievability of
$\frac{1}{n}I(M;Y_i^n |E=k)$ depends on the image-size
characterization, which on $A_k$ is the same as the achievable entropy
characterization.  


\end{document}